\documentclass[superscriptaddress,showpacs,aps,floats,twocolumn,prb]{revtex4}

\usepackage{graphicx}

\begin{document}

\newcommand{\LBCO}{La$_{2-x}$Ba$_x$CuO$_4$}
\newcommand{\LSCO}{La$_{2-x}$Sr$_x$CuO$_4$}
\newcommand{\YBCO}{YBa$_{2}$Cu$_3$O$_{7-\delta }$}
\newcommand{\BSCCO}{Bi$_2$Sr$_2$CaCu$_2$O$_8$}

\title{High resolution X-ray scattering studies of structural phase transitions in underdoped \LBCO}


\author{Y. Zhao}
\affiliation{Department of Physics and Astronomy, McMaster University, Hamilton, Ontario, L8S 4M1, Canada.}

\author{B. D. Gaulin}
\affiliation{Department of Physics and Astronomy, McMaster University, Hamilton, Ontario, L8S 4M1, Canada.}
\affiliation{Canadian Institute for Advanced Research, 180 Dundas St. W., Toronto, Ontario, M5G 1Z8, Canada}

\author{J. P. Castellan}
\affiliation{Department of Physics and Astronomy, McMaster University, Hamilton, Ontario, L8S 4M1, Canada.}

\author{J. P. C. Ruff}
\affiliation{Department of Physics and Astronomy, McMaster University, Hamilton, Ontario, L8S 4M1, Canada.}

\author{S. R. Dunsiger}
\affiliation{Department of Physics and Astronomy, McMaster University, Hamilton, Ontario, L8S 4M1, Canada.}

\author{G. D. Gu}
\affiliation{Physics Department, Brookhaven National Laboratory, Upton, New York 11973, USA}

\author{H. A. Dabkowska}
\affiliation{Department of Physics and Astronomy, McMaster University, Hamilton, Ontario, L8S 4M1, Canada.}

\begin{abstract}

We have studied structural phase transitions in high quality underdoped \LBCO\ single crystals using high resolution x-ray 
scattering techniques.  Critical properties associated with the continuous High Temperature Tetragonal (HTT, $I4/mmm$) to Middle Temperature Orthorhombic (MTO, $Cmca$) phase transition were investigated in single crystal samples 
with x=0.125, 0.095, and 0.08 and we find that all behavior is consistent with three dimensional XY criticality, as expected from
theory.  Power law behavior in the orthorhombic strain, 2(a-b)/(a+b), is observed over a remarkably wide temperature range, spanning most of the MTO regime in the phase diagram.  Low temperature measurements investigating the Low Temperature  Tetragonal (LTT, $P4_{2}/ncm$) phase, below the strongly discontinuous MTO$\to$LTT phase transition, in x=0.125 and x=0.095 samples show that the LTT phase is characterized by relatively broad Bragg scattering, compared with that
observed at related wavevectors in the HTT phase.  This shows that the LTT phase is either an admixture of tetragonal and orthorhombic phases, or that it is orthorhombic with very small orthorhombic strain, consistent with the ``less orthorhombic" low temperature structure previously reported in mixed La$_{2-x}$Sr$_{x-y}$Ba$_y$CuO$_4$ single crystals.  We compare the complex temperature-composition phase diagram for the location of structural and superconducting phase transitions in underdoped \LBCO\ and find good agreement with results obtained on polycrystalline samples.

\end{abstract}

\date{\today }


\maketitle

\section{\label{} Introduction}

The complex interplay between spin, charge, and lattice degrees of freedom in the quasi-two dimensional copper-oxide 
high temperature superconductors have been the subject of intense interest since the discovery of
superconductivity in the \LBCO\ system some 21 years ago\cite{Bednorz:1986}.  Both \LBCO\ and \LSCO\ display a fascinating series of structural, magnetic and superconducting phase transitions as a function of temperature\cite{Kastner:1998}.  While \LBCO\ was the first layered cuprate high T$_{c}$ superconductor to be discovered, difficulties associated with the 
growth of high quality single crystals have significantly limited its study.  As a result the \LBCO\ family is much less studied than the \LSCO\ family and other high temperature superconductors which have an extended history of being grown and characterized in single crystal form\cite{Kastner:1998}, such as the \YBCO\ and \BSCCO\ families\cite{Birgeneau:2006,Eschrig:2006,Fong:1999,Castellan:2006}.  

Recently, significant progress has been made in growing the \LBCO\ family of materials in single crystal form, and  this has enabled several important new studies of this and related systems\cite{Fujita:2004,Reznik:2006,Tranquada:2004,Kimura:2005}.  It is therefore timely to perform high resolution structural studies of these new single crystals, and to compare to previous studies on \LBCO\ in polycrystalline form\cite{Suzuki:1989,Suzuki:1989a}.

One of the many interesting properties of the \LBCO\ family is the sequence of structural phase transitions which this material displays on cooling 
below room temperature for underdoped Ba concentrations (x$\lesssim$ 0.18). Previous studies on polycrystalline \LBCO\ shows three different structures, which proceed from High Temperature Tetragonal (HTT, $I4/mmm$), to Middle Temperature Orthorhombic (MTO, $Cmca$) and finally to Low Temperature Tetragonal (LTT, $P4_{2}/ncm$)\cite{Axe:1989,Axe:1989a,Suzuki:1989,Suzuki:1989a,Adachi:2001}. The HTT$\to$MTO and the MTO$\to$LTT phase transition temperatures are referred to as $T_{d1}$ and $T_{d2}$, respectively.   The HTT$\to$MTO transition is 
continuous, while the MTO$\to$LTT transition is known to be strongly discontinuous.  These structures are closely 
related to the magnetic and electronic properties of the \LBCO\ and \LSCO\ families.  The phase diagram of the \LBCO\ system contains a dome
of LTT phase, which is centred around x=0.125.  This Ba-concentration corresponds to a steep depression of the superconducting $T_C$ as a function of concentration, known as the 1/8 anomaly\cite{Axe:1989a,Moodenbaugh:1988}.  The \LSCO\ system shows a much smaller $\sim$ 10$\%$ dip in $T_C$ at x=0.125 and the absence of the LTT phase at low temperatures\cite{Nagano:1993,Radaelli:1994}.  The 1/8 anomaly within the LTT phase also corresponds to strong
incommensurate magnetic long range order at temperatures just below the completion of the MTO-LTT phase transition\cite{Fujita:2004,Tranquada:2004}.  Clearly, the structural,
magnetic, and superconducting properties of the \LBCO\ and \LSCO\ systems are strongly coupled.

The critical phenomena associated with the HTT-MTO transition has been previously studied in pure La$_{2}$CuO$_{4}$ as well as in \LSCO\ in single crystal and polycrystal form\cite{Birgeneau:1987,Vaknin:1987,Boni:1988,Braden:1994,Ting:1993,Thurston:1989}, as single crystals of these materials have existed for some time.  These
studies show the HTT$\to$MTO phase transition to be characterized with an order parameter critical exponent $\beta$ varying from 0.28 to 0.37\cite{Birgeneau:1987,Vaknin:1987,Boni:1988,Braden:1994,Ting:1993,Thurston:1989}. Studies on polycrystalline samples of \LBCO\ by Susuki et al
produced estimates for $\beta$ $\sim$ 0.33\cite{Suzuki:1989,Suzuki:1989a}, and which are consistent with expectations for 3D universality\cite{Collins:1989}.

In this paper, we report the successful growth of large \LBCO\ single crystals with x=0.095 and 0.08, and a high resolution x-ray diffraction study on the x=0.125, 0.095 and  0.08 compounds in this family. This study focusses on a comparison between the structural and superconducting phase diagrams in polycrystalline and single crystal materials, critical phenomena associated with the HTT$\to$MTO phase transition, and the nature of the LTT phase in x=0.125 and 0.095 samples at low temperatures.

\section{\label{} Experiment details}
\subsection{\label{} Crystal Growth}

\begin{figure}
\includegraphics[width=0.9\columnwidth]{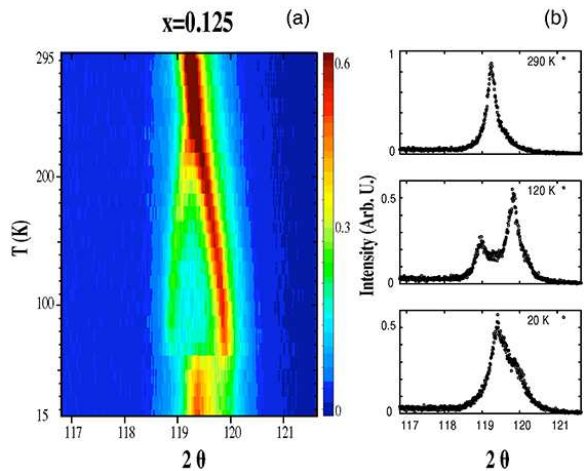}
\caption {(a), High resolution longitudinal scans of the (3, 3, 0)$_{HTT}$ Bragg peak in single crystal \LBCO, x=0.125 are shown as a function of temperature. (b) Representative longitudinal scans at T=290 K, 120 K, and 20 K from which the color contour map in (a) was made.}  
\end{figure}

\begin{figure}
\includegraphics[width=0.9\columnwidth]{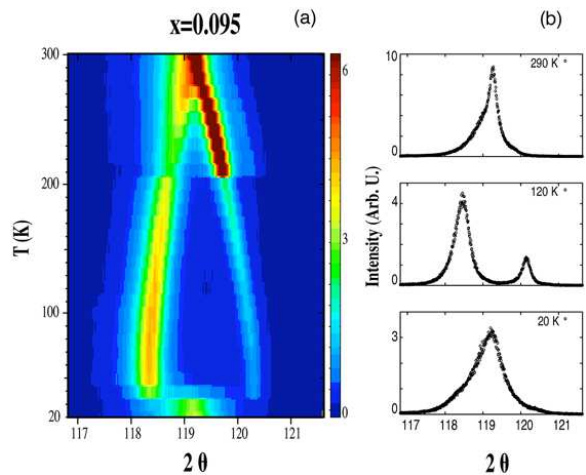}
\caption {(a), High resolution longitudinal scans of the (3, 3, 0)$_{HTT}$ Bragg peak in single crystal \LBCO, x=0.095 are shown as a function of temperature. (b) Representative longitudinal scans at T=290 K, 120 K, and 20 K from which the color contour map in (a) was made.}  
\end{figure}

We studied three high quality \LBCO\ single crystals with x=0.125, 0.095 and 0.08.  All crystals were grown by using traveling solvent, floating zone image furnace techniques. The x=0.125 sample was grown separately, and the details of this growth have been previously discussed\cite{Fujita:2004,Tranquada:2004}.

The x=0.095 and 0.08 \LBCO\ single crystal growths followed similar processess, and employed polycrystalline La$_{2}$O$_{3}$, BaCO$_{3}$ and CuO as starting 
materials to make the initial, polycrystalline feed rod and solvent. For the production of the feed rods, the starting materials were mixed to
give an initial ratio of La:Ba:Cu=1.875:0.125:1. These materials were mixed, ground, and annealed at 980$^{\circ}$C for 12 hours 
in air. This process was repeated twice in order to ensure homogeneous feed rods.  To compensate for Cu evaporation during the crystal growth, the pre-annealed feed rods were mixed with extra CuO. A further 1\% and 2\% mol CuO was added to the starting polycrystalline materials and thoroughly mixed to prepare the two final feed rods, respectively. The final feed rods were heated to a temperature of 1190$^{\circ}$C, at a rate of 100$^{\circ}$C/hour.  They were held at this temperature for 12 hours.  We also employed a solvent, formed from the original polycrystalline feed rod, with CuO added so as to reach a final ratio of constituent atoms (La$_{1.875}$Ba$_{0.125}$):Cu=3:7. After mixing and sintering, small disks weighing $\sim$ 0.44 g were cut out and used as solvents in the subsequent single crystal growths.

The single crystal growths were carried out using a four-mirror image furnace (Crystal System Inc.). A small pure La$_{2}$CuO$_{4}$ single crystal was employed as the seed rod for both growths. The growths were carried out in an O$_2$ atmosphere at pressures of 165 kPa and 182 kPa for the two crystal growths.  The growth rate was 1mm/h with a counter-rotation speed of 25 rpm for feed and seed rods for both growths.

Upon completion of the growths, the as-grown single crystals were kept above 100$^{\circ}$C in a furnace to prevent hydrolysis of the 
material, which is known to be problematic for single crystal \LBCO. The two crystals, which are identified in this study as being at x=0.095 and 0.08, were of almost identical dimensions of 80 mm long by  5 mm in diameter as-grown. Within the first week following completion of the growths, the initial $\sim$ 30 mm of the crystals turned to dust as a result of hydrolysis of the second phase. The undamaged part of both crystals was stable.  They had approximate dimensions of 50 mm long by 5 mm in diameter for x=0.095 and 55 mm long by 5 mm in diameter for x=0.08.
These volumes are sufficiently large for advanced characterization by neutron scattering techniques, and indeed a program of neutron measurements has been carried out on these samples\cite{Dunsiger:2007}.  

We note that while the two crystal growths were initiated with similar starting materials, and the growths followed similar procedures, the Ba/La ratio, as identified by T$_{d1}$ and T$_{d2}$, were different at the $\sim$ 15$\%$ level. This originates from Cu evaporation during the growth. All the phase transitions observed (structural, magnetic, and superconducting) are nevertheless very sharp in temperature, indicating excellent homogeneity of concentration within the individual single crystals.

\subsection{\label{} X-ray diffraction}

Single crystal samples with approximate dimensions 8 mm$\times$8 mm$\times$1 mm for x=0.125, and 5 mm$\times$5 mm$\times$1 mm for x=0.095 and 0.08, were cut from large single crystals of \LBCO.  These were sequentially attached to the cold finger of a closed cycle refrigerator and mounted within a four circle x-ray 
diffractometer.  Cu K$_{\alpha 1}$ radiation from an 18kW rotating anode x-ray generator was selected using a perfect Germanium (111) single crystal monochrometer. A Bruker Hi-Star multi-wire area detector was placed on the detector arm, 76 cm from the sample allowing an angular resolution of approximately  0.01 degrees to be achieved. All measurements focused on (3, 3, 0)$_{HTT}$ Bragg peak of the samples, using notation appropriate to the high temperature tetragonal phase. As we were interested in critical phenomena, the sample was mounted in a Be can and in the presence of a helium exchange gas and the sample temperature was stabilized to $\sim$ 0.005 K for all measurements. 

\section{\label{} Experimental Results}

\subsection{\label{} Identification and Nature of Phases}

\begin{figure}
\includegraphics[width=0.9\columnwidth]{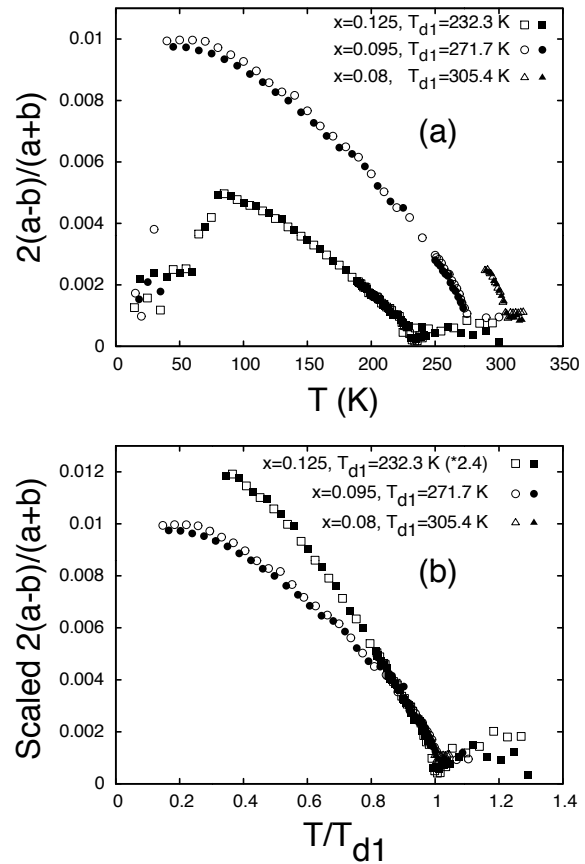}
\caption {(a) The orthorhombic strain vs. temperature is plotted for \LBCO\ x=0.125, 0.095 and 0.08 single crystal samples. The open and filled symbols represent warming and cooling cycles, respectively.  The orthorhombic strain is obtained by fitting longitudinal scans, shown in Figs. 1 and 2.  (b) The same orthorhombic strain vs. temperature as in (a) but now plotted vs T/T$_{d1}$ and the strain has been scaled (for the x=0.125 sample, by a factor of 2.4) to emphasize universal behavior for T/T$_{d1}$ greater than 0.8. }  
\end{figure}

Two dimensional maps of the scattering around the (3, 3, 0)$_{HTT}$ Bragg peaks of all three x=0.125, 0.095 and 0.08 \LBCO\ samples were acquired as a function of temperature.  Each data set consisted of a sample angle rock through the Bragg peak which was integrated in the vertical 
direction and plotted as a function of scattering angle, 2$\theta$.  A longitudinal cut through this two dimensional data set was performed, giving rise to the longitudinal scans shown in Fig. 1b for the x=0.125 sample, and Fig. 2b for the x=0.095 sample.  Similar data sets taken over a more restricted temperature regime for the x=0.08 sample are of similar quality, but are not shown.

These data sets can be put together to display the full temperature dependence of the longitudinal scans, and this is what is shown in Figs. 1a and 2a for the x=0.125 and x=0.095 samples, respectively.  These data sets clearly show the bifurcation of a single Bragg peak into two, and then back into one, as the temperature is decreased from room temperature to 20 K, signifying the sequence of phase transitions HTT$\to$MTO$\to$LTT.   The fact that two Bragg features can be seen in a single longitudinal scan within the MTO phase is indicative of twinning within the orthorhombic phase, although the two twin domains which are observed do not possess equal volume fraction within the crystal;  one Bragg feature is considerably stronger in intensity than the other.  A minority and majority twin domain is clearly present, but the relevant volume fraction can change from one thermal cycle to the next.  For example, the x=0.095 data set shown in Fig. 2(a) shows data from two independent thermal cycles, one ending with a lowest temperature of $\sim$ 200 K, while the next beginning a new thermal cycle at 200 K.  In the first of these, the high angle Bragg peak is the majority domain, while in the second cycle, the lower angle Bragg peak is the majority domain.

\begin{figure}
\includegraphics[width=0.9\columnwidth]{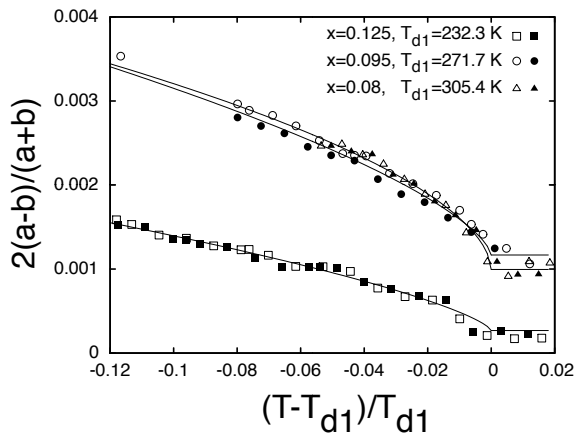}
\caption {The orthorhombic strain is plotted vs reduced temperature, (T-T$_{d1}$/T$_{d1}$) for the x=0.125, 0.095, and 0.08 \LBCO\ samples at small values of reduced temperature, near T$_{d1}$.  The open and filled symbols show data from warming and cooling cycles, respectively.  Fits of the data to the form of the order parameter squared vs reduced temperature, Eq. 1, 
used to extract values of $\beta$ are shown as the solid lines.}  
\end{figure}

The fact that we observe both twin domains in the MTO phase means that the peak positions, the lattice parameters, and consequently the
orthorhombic strain, 2(a-b)/(a+b), can be determined as a function of temperature.  This is shown for all three samples in Fig. 3a.  The single (3, 3, 0)$_{HTT}$ Bragg peak breaks into (6, 0, 0)$_{MTO}$ and (0, 6, 0)$_{MTO}$ near T$_{d1}$=232 K, 272 K and 305 K in the x=0.125, 0.095, and 0.08 samples, respectively, before reforming into a single (3, 3, 0)$_{LTT}$ Bragg peak near T$_{d2}$=60 K, 45 K, and 35 K, respectively.

Examination of Figs 1-3 shows two qualitative features of the evolving structures.  Note that for ease of comparison, the 2$\theta$ range of the
scattering in Figs. 1 and 2 is the same.  First, the orthorhomic strain decreases quite substantially with increasing Ba concentration.  The lowest
temperature strain, for example, in the  x=0.125 sample is roughly half that of the x=0.095 sample.  Secondly and more importantly, the longitudinal profile of the (3, 3, 0)$_{LTT}$ peak at the lowest temperatures measured, well within the LTT phase, is considerably broader than the
corresponding profile of (3, 3, 0)$_{HTT}$.  This is true for both the x=0.125 sample and the x=0.095 sample as can be seen by comparing the top and bottom panels of Fig. 1b (for x=0.125) and Fig. 2b (for x=0.095).  This shows that the LTT phase is either an admixture of a tetragonal and an orthorhombic phase, as was suggested by electron microscopy on 
an earlier generation of \LBCO\ crystals\cite{Zhu:1994}, or that it is itself othorhombic with a very small orthorhombic strain.  In either case it is not as ``tetragonal" as the HTT phase, and is consistent with the ``less orthorhombic" low temperature structures proposed previously for La$_{2-x}$Sr$_{x-y}$Ba$_y$CuO$_4$ single crystals\cite{Fujita:2002b}.

\subsection{\label{} Critical Phenomena at the HTT$\to$MTO Phase Transition}

Longitudinal scans of the form shown in Fig. 1b and 2b were fit for the purpose of extracting the peak positions in 2$\theta$ and 
therefore the d spacings associated with the MTO phase.  This is straightforward for data far removed from the HTT$\to$MTO phase transition, as the two peaks are well defined and separated, as can be seen in the middle panels of Fig. 1b and 2b.  Closer to the phase transition, one peak may appear as a shoulder to the other, and it is more difficult to ascribe unique values to the two lattice parameters.  We fit these data in two different ways in order to attain robust values for the lattice parameters close to the transition.  One of these was to simply fit the longitudinal scans to sums of Lorentzians or Lorentzians raised to an adjustable exponent, while a second technique was to look for zeros in the derivatives of the intensity as a function of 2$\theta$.  These gave consistent results for the lattice parameters, giving us confidence that the orthorhombic strain could be estimated accurately close to the transition. However, this technique also gives non-zero values for the orthorhombic strain, albeit relatively small ones, within the HTT phase.  

Previous work on the HTT$\to$MTO phase transition in polycrystalline \LSCO\ and \LBCO\ samples show the 
orthorhombic strain to scale as the square of the order parameter\cite{Birgeneau:1987,Boni:1988,Suzuki:1989,Suzuki:1989a}.  Consequently we examined 
the critical behaviour of the orthorhombic strain in our \LBCO\ single crystals by fitting the measured strain as a function of temperature to: 
\begin{equation}
\label{ }
\Delta=\Delta_{0}\times  (\frac{T_{d1}-T} {T_{d1}})^{2\beta}+Background
\end{equation}
where the square of the order parameter, $\Delta$, is the orthorhombic strain, 2(a-b)/(a+b), and the background accounts for finite strain within the HTT phase introduced by the fitting process described above. The results of this fitting is shown in Fig. 4, which shows the orthorhombic strain as 
a function of reduced temperature, (T-T$_{d1}$)/T$_{d1}$, in the region of small reduced temperature close to T$_{d1}$.  Clearly this description of the data is very good.  It results in accurate estimates for both $\beta$ and T$_{d1}$.  These are T$_{d1}$=232.3 $\pm$ 0.7 K, 271.7 $\pm$ 1 K, and 305.4 K $\pm$ 1 K for the x=0.125, 0.095, and 0.08 samples, respectively.  The extracted values for $\beta$ are 0.35 $\pm$ 0.03, 0.34 $\pm$ 0.04 and 0.28 $\pm$ 0.06, respectively. 

Using these values of T$_{d1}$ for each of the three samples, we can scale the plot of orthorhombic strain vs temperature, Fig. 3a, so as to give scaled orthorhombic strain vs T/T$_{d1}$, which is shown in Fig. 3b.  We see that above T/T$_{d1}$ $\sim$ 0.8 the orthorhombic strains for all three samples collapse to a single curve.  We therefore expect universal behaviour in this regime, which is borne out by the similarity in the extracted values for the critical exponent $\beta$ at all three Ba concentrations.

\begin{figure}
\includegraphics[width=0.9\columnwidth]{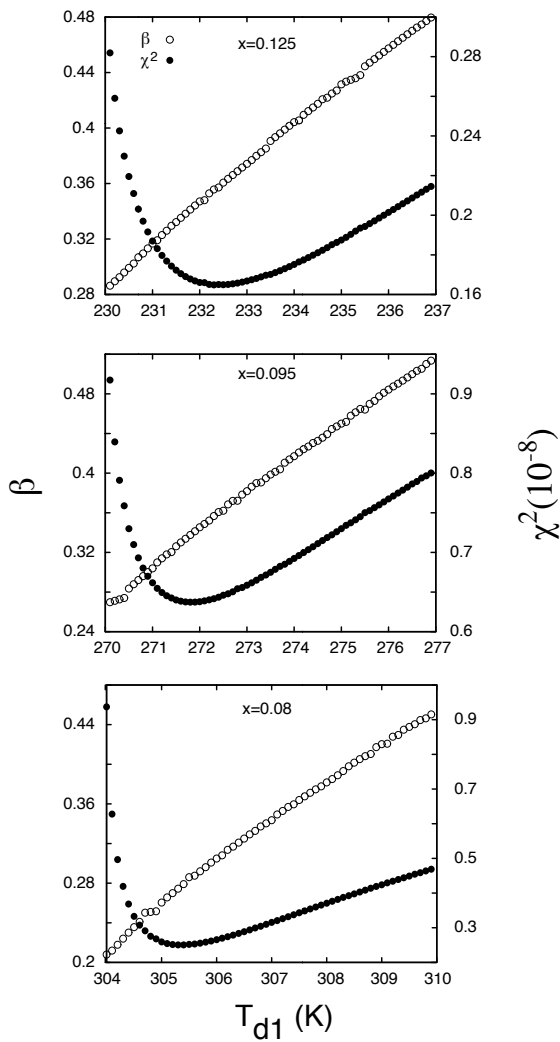}
\caption {The dependence of critical exponent $\beta$ and goodness-of-fit parameter $\chi^2$ are shown as a function  of the assumed value of T$_{d1}$ for x=0.125 (upper panel), x=0.095 (middle panel) and x=0.08 (lower panel) \LBCO\ single crystal samples.  The uncertainty in $\beta$ is largely determined by the uncertainty in critical temperature T$_{d1}$.}  
\end{figure}

The uncertainties associated with the critical exponent $\beta$ are largely determined by the uncertainties in the critical temperature, T$_{d1}$, derived from the fits to the critical behaviour.  We performed fits to Eq. 1 using T$_{d1}$ set to a range of values around the approximate phase transition temperature, and then allowed the fit to adjust the other parameters in Eq. 1.  This gives a monotonically increasing estimate for $\beta$ as a function of increasing T$_{d1}$.  Best estimates for $\beta$ and T$_{d1}$ are given by the minimum in the goodness-of-fit parameter $\chi^2$ which we define as:
\begin{equation}
\label{ }
\chi^{2}=\frac{\sum(\Delta_{measured}-\Delta_{calculated})^{2}} {N}
\end{equation}
where N is the number of data points.

$\beta$ and $\chi^2$ are shown as a function of T$_{d1}$ for the x=0.125 (top panel), x=0.095 (middle panel), and x=0.08 (bottom panel) samples in Fig. 5.  The uncertainty in $\beta$ is determined by the corresponding uncertainty in T$_{d1}$, and it is roughly 10$\%$ for the x=0.125 and 0.095 samples where we have an extended data set throughout the MTO phase, and roughly 20$\%$ for the x=0.08 sample where the data set is restricted to temperatures close to T$_{d1}$.

\begin{figure}
\includegraphics[width=0.9\columnwidth]{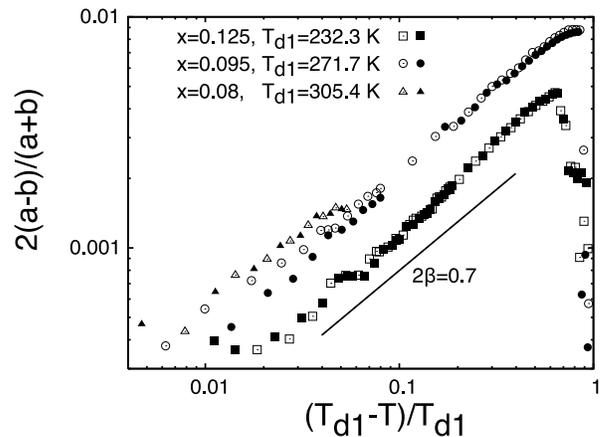}
\caption {The orthorhombic strain is plotted as a function of reduced temperature, (T$_{d1}$-T)/T$_{d1}$, on a log-log scale for the x=0.125, 0.095 and 0.08 single crystal \LBCO\ samples.  The open and filled symbols show results from warming and cooling cycles, respectively.  For comparison power law behavior showing $\beta$=0.35, indicative of the theoretically expected 3D XY universality class, is indicated as the straight line on this log-log plot.}  
\end{figure}

Investigation of the critical properties at the HTT$\to$MTO phase transition in polycrystaline \LSCO\cite{Birgeneau:1987,Boni:1988}\ and \LBCO\cite{Suzuki:1989,Suzuki:1989a}\ anticipated 3D XY universality on the basis of a Landau expansion appropriate to this ferroelastic system.  These early results on polycrystalline systems were consistent with the $\beta$=0.35 expected from 3D XY universality\cite{Le-Guillou:1977,Le-Guillou:1980}.  However, these earlier estimates for $\beta$ spanned the range from 0.28 to 0.37, ignoring uncertainties associated with the estimates, which covers all standard 3D universality classes: Heisenberg ($\sim$0.37), XY ($\sim$0.35), Ising ($\sim$0.32) and which begins to approach values consistent with tricritical phenomena (0.25)\cite{Collins:1989}.

Figure 6 shows the orthorhombic strain, 2(a-b)/(a+b), plotted as a function of the reduced temperature, (T$_{d1}$-T)/T$_{d1}$, on a log-log plot 
in order to identify the expected power law regime.  For comparison a straight line appropriate to $\beta$=0.35 and 3D XY universality is also plotted.  For each sample, two data sets are plotted, one for a warming run and one for a cooling run.  We observe very similar power law behaviour in all three samples, and behaviour which is very much consistent with 3D XY universality as anticipated theoretically.  We also see, at least for the x=0.125 and 0.095 samples for which we have data over the entire MTO phase regime in temperature, that a single power law is a remarkably good descriptor of the data over a very large temperature regime.  There appears to be a slight increase in slope for reduced temperatures greater than $\sim$ 0.2, but overall, power law-like growth of the orthorhombic strain is observed over almost two decades in reduced temperature.  This is in contrast to most critical phenomena, wherein asymptotic critical behaviour is expected to cross over to a mean field-like regime, as one moves away from the critical temperature.

Taken together our orthorhombic strain measurements show critical behaviour at the HTT$\to$MTO phase transition in single crystal \LBCO\ over a broad range of concentration which is characterized by $\beta$=0.34 $\pm$ 0.04.  This result clearly demonstrates 3D universality, and is consistent with 3D XY universality which is expected based on Landau theory.  It is also largely consistent with previous experimental work on single crystal and polycrystal \LSCO\ and polycrystalline \LBCO, much of which centred on measurements of superlattice Bragg peak intensities within the MTO structure, as opposed to measurements of the orthorhombic strains\cite{Braden:1994,Thurston:1989}.  Superlattice Bragg peak intensities near continuous phase transitions can be difficult to interpret, as they can be influenced by extinction and by fluctuations above the phase transition.  This latter effect manifests itself in upwards curvature and difficulty identifying a precise phase transition temperature, which in turn can lead to uncertainty in critical exponents. 

\begin{table}
\caption{\label{tab:table1} Summary of structural and superconducting phase transition temperatures in single crystal \LBCO}
\begin{ruledtabular}
\begin{tabular}{ccccc}
x&$T_{d1}$(K)&$T_{d2}$(K)&$T_{c}$(K) & $\beta$\\
\hline
0.125 & 232.3 & 60 & 4\footnotemark[1] & 0.35 $\pm$ 0.03\\
0.095 & 271.7& 45\footnotemark[2] & 27\footnotemark[2] & 0.34 $\pm$ 0.04 \\
0.08& 305.4 & 35\footnotemark[2] & 29\footnotemark[2] & 0.28 $\pm$ 0.06 \\
\end{tabular}
\end{ruledtabular}
\footnotetext[1]{From Ref.~\onlinecite{Fujita:2004}.}
\footnotetext[2]{From Ref.~\onlinecite{Dunsiger:2007}.}
\end{table}

\subsection{\label{} Phase Diagram and Comparison to Polycrystalline Materials}

It is of interest to compare the \LBCO\ phase diagram known to characterize pre-existing polycrystalline samples with that determined for the high quality single crystals
in the present studies.  A rather detailed comparison can be carried out, as two structural and one superconducting transition temperature characterize \LBCO\ samples in this underdoped concentration range.  The phase transitions measured for the single crystals in this study are summarized in Table 1.  The critical exponent $\beta$
relevant to the HTT$\to$MTO structural transition is also shown in the same table for reference.

The superconducting transition temperatures were determined from SQUID magnetometry as reported by Dunsiger et al.\cite{Dunsiger:2007} for the x=0.095 and x=0.08 samples, and 
by Fujita et al.\cite{Fujita:2004} for the x=0.125 sample.  The strongly first order  MTO$\to$LTT transition is measured both by the abrupt change in the orthorhombic strain seen in Fig. 1 and 2, for the x=0.125 and x=0.095  samples, respectively, as well as by the appearance of the (0, 1, 0) superlattice Bragg peak intensity as again reported by Dunsiger et al.\cite{Dunsiger:2007} for the x=0.095 and x=0.08 samples, and by Fujita et al. for the x=0.125 sample\cite{Fujita:2004}. 

\begin{figure}
\includegraphics[width=0.9\columnwidth]{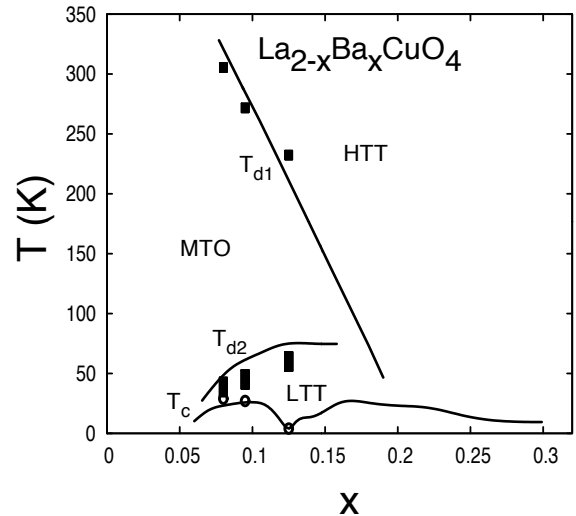}
\caption {Phase boundaries identifying structural and superconducting phases of \LBCO\ single crystals are plotted on the phase diagram derived from 
previously studied polycrystalline samples.  The structural transitions at T$_{d1}$ and T$_{d2}$ are indicated by filled squares, while superconducting T$_C$'s are indicated
by open circles.  The first order transition at T$_{d2}$ is indicated by a bar $\sim$ 10 K wide, showing the onset to completion of the phase transition.  Solid lines showing phase boundaries from polycrystalline \LBCO\ are taken from Adachi et al.\cite{Adachi:2001} }
\end{figure}

Figure 7 shows the \LBCO\ phase diagram with HTT, MTO, and LTT phases indicated. The HTT$\to$MTO and MTO$\to$LTT transitions are shown as filled squares for the three Ba concentrations measured.  The discontinuous transition at T$_{d2}$ is indicated as a bar, in order to show the onset to completion of the transition, which is $\sim$ 10 K wide.  T$_{d2}$ in Table 1 is the midpoint of the transition.  The superconducting transitions are given by the open circles, and they indicate the onset of the superconductivity, which is also what is listed in Table 1.  Previous results for these same phase boundaries as determined for polycrystalline \LBCO\ samples are shown as the solid lines in Fig. 7.  These results were extracted from Adachi et al.\cite{Adachi:2001}  and are reproduced here.

As can be seen on inspection of Fig. 7, the agreement between the structural and superconducting phase boundaries in polycrystalline \LBCO\ and the new floating zone image furnace grown single crystals is remarkably good.  The absolute values for T$_{d2}$ are systematically high, at the 10$\%$ level for the polycrystalline materials as compared to the single crystals, but overall the full level of agreement is excellent.  In particular we see that good agreement between the two for T$_{d1}$ means that this transition can be used as an accurate marker for the Ba concentration in single crystal \LBCO, as T$_{d1}$ has such strong Ba dependence.  The image furnace single crystals were grown without crucibles, and are expected to be of higher purity than the corresponding polycrystalline materials grown from a flux melt in a crucible.  The similarity between the overall phase diagrams in polycrystalline and image furnace grown single crystal \LBCO, implies an insensitivity of these phase boundaries to this level of imperfection.

\section{\label{} Conclusions}

We have successfully grown large single crystals of \LBCO\ with x=0.095 and 0.08 using floating zone image furnace techniques.  These single crystals are sufficiently large so as to enable neutron scattering studies, which will be reported separately\cite{Dunsiger:2007}.  High resolution single crystal x-ray diffraction measurements were carried out on these samples, as well as on a high quality x=0.125 single crystal.  These measurements focus on the (3, 3, 0)$_{HTT}$ Bragg peak and show the HTT$\to$MTO$\to$LTT sequence of structural phase transitions known to be relevant to underdoped \LBCO.  The measurements also clearly show anomolous longitudinal broadening of the (3, 3, 0)$_{LTT}$ Bragg peaks in the x=0.095 and x=0.125 samples at low temperatures, indicating that the LTT phase is not a simple tetragonal phase, but rather an admixture of tetragonal and orthorhombic phases, or an orthorhombic phase with very small orthorhombic strain.  Critical
orthorhombic strain measurements near the continuous HTT$\to$MTO phase boundary show clear 3D universality, with universal behavior observed in the orthorhombic strain vs T/T$_{d1}$ for the three x=0.125, 0.095 and 0.08 samples.  The best estimate for a common critical exponent $\beta$ for these samples is $\beta$=0.34 $\pm$ 0.04, which is consistent with 3D XY universality expected theoertically for such ferroelastic transitions.  A detailed comparison of the \LBCO\ phase diagram incorporating structural and superconding phase boundaries at this underdoped concentration regime indicates excellent agreement with pre-existing data based on polycrystalline samples.

It is a pleasure to acknowledge the contributions of Ms. Ann Kallin to the single crystal growth.  This work was supported by NSERC of Canada.  Gu was supported by the US Department of Energy under contract number DE-AC02-98CH10886.

\end{document}